\documentclass[aps,prd,epsfig,preprint,amsmath,amssymb,superscriptaddress,nofootinbib]{revtex4}
\usepackage{multirow}
\usepackage{graphicx,bm}
\usepackage[usenames]{color}
\usepackage{float}
\usepackage{caption}
\usepackage{subcaption}
\usepackage{slashed}
\begin{document}
\draft
\title{Constraints on the anomalous Higgs boson couplings in $Z\gamma\gamma$ production at muon collider}

\author{S. Spor}
\email[]{serdar.spor@beun.edu.tr}
\affiliation{Department of Medical Imaging Techniques, Zonguldak B\"{u}lent Ecevit University, 67100, Zonguldak, T\"{u}rkiye.}

\begin{abstract}

We investigate the sensitivities on the Wilson coefficients of dimension-six operators associated with the anomalous $H\gamma\gamma$, $H\gamma Z$ and $HZZ$ vertices through the process $\mu^- \mu^+ \rightarrow Z\gamma\gamma$ at future muon collider. Signal events involving Higgs-gauge boson interactions and relevant backgrounds events at the muon collider designed with a center-of-mass energy of 10 TeV and an integrated luminosity of 10 ab$^{-1}$ are generated in Madgraph within the framework of the model-independent Standard Model effective field theory. Realistic detector effects of the muon collider detector are simulated by Delphes. The limits at 95\% C.L., $3\sigma$ and $5\sigma$ on the coefficients $\overline{c}_{HB}$ and $\overline{c}_{HW}$ are obtained without and with systematic uncertainty of 15\% and compared with the present experimental and phenomenological limits.

\end{abstract}

%\keywords{Models Beyond the Standard Model, Compact Linear Collider, Muon Collider, Higgs Boson Couplings.}

\maketitle

\section{Introduction} \label{Sec1}

With the discovery of the Higgs particle \cite{Aad:2012les,Chatrchyan:2012les}, intensive studies \cite{Aad:2022owv,Tumasyan:2022uez} have been done to reveal the properties of the Higgs particle at colliders, both through the Standard Model (SM) predictions and scenarios beyond the SM. Despite the discovery of the Higgs boson, the Higgs sector's structure is still unclear. Studies \cite{Aad:2016saf,Aad:2020czq,Aad:2020jkn,Sirunyan:2020wxc,Tumasyan:2022ssx} have shown that the discovered Higgs boson is a CP-even scalar with CP-conserving interactions and is consistent with SM predictions. However, the SM is not perfect and there is great motivation to search for new physics beyond the SM to address many unsolved questions.

In particle physics, one crucial area of study for future experiments is the search for CP violation. Future high-energy physics experiments have special qualifications to test for CP violation in the Higgs boson interactions. Among future collider designs, the multi-TeV muon collider incorporates many of these special features. Owing to their greater mass compared to electrons, muons emit about two billion times less synchrotron radiation, which facilitates their acceleration to high energies within a circular collider. Furthermore, as fundamental leptons rather than composite particles like protons, muons deposit the entirety of their center-of-mass energy into the collision, enhancing the effective energy reach. In addition, lepton colliders inherently offer a much cleaner experimental environment and substantially reduced background levels relative to hadron colliders, thereby improving the precision of measurements and the sensitivity to new physics phenomena. Since CP violation impacts are negligible in the SM, they offer great testing opportunities to compare the performance of future facilities. The search for precise measurements of Higgs boson couplings at multi-TeV muon colliders has recently become more important in the study of physics beyond the SM \cite{Chiesa:2020jer,Han:2021opj,Buttazzo:2021uha,Han:2021egh,Forslund:2022hjw,Celada:2024tgv,Chiesa:2024asx,Cassidy:2024uhn,Dermisek:2024qaz,Spor:2024ghq,Forslund:2024yhz,Gurkanli:2025wcv,Chen:2022ygc,Li:2024pmz,Spor:2024yln}. The presence of theories beyond the SM that can predict effects of CP violation will have significant outcomes for particle physics in the future.

In this study, we use the Standard Model Effective Field Theory (SMEFT) as a model-independent way to parameterize the effects of new physics at the high energies of the 10 TeV muon collider to improve the search for CP-violating phenomena. Any extra CP-violating sources identified by the new physics at high energies can be expressed in terms of CP-violating SMEFT parameters in this framework \cite{Degrande:2022mcx}.

\section{Effective operators and Higgs-gauge boson interactions} \label{Sec2}

A new physics model with additional fields and interactions is needed to make progress in particle physics. In order to understand the new physics in light of experimental data, it is necessary to establish the relationship between the observables at the electroweak scale and physics beyond the SM at a relatively high scale. In order to describe new physics at high mass scales, effective field theory (EFT) presents a model-independent way. EFTs are a tool to bridge the gap between scales and reveal the indirect new physics effects from experimental data. The EFT is only valid up to a particular energy scale ($\Lambda$); in order for the theory to be practical, it must be above the energy scale ($E$) that is immediately accessible experimentally. In other words, when $\Lambda \gg E$, the EFT gives a good approximation. 

One of the most effective ways to measure the effects of the new physics is to use higher-dimensional effective operators consisting of SM fields that obey the SM gauge symmetry. These are named the SMEFT operators. In this approximation, additional dimension-six operators are added to the SM Lagrangian to ensure conservation of lepton and baryon numbers, and dimension-eight and higher operators are neglected for further suppression. The SMEFT operators can be expressed in a theoretical context in the Strongly Interacting Light Higgs (SILH) basis. In the EFT framework on the SILH basis, the effects of unobserved states, which are predicted to arise at energies bigger than an effective scale identified with the $W$-boson mass $m_W$ or with the vacuum expectation value of the Higgs field $\nu$, are parameterized using high-dimensional operators for interactions beyond the SM.

An effective theory of a light composite Higgs boson, which is responsible for electroweak symmetry breaking and arises as the pseudo-Goldstone boson from a strongly interacting sector, reveals the SILH basis \cite{Tosciri:2021yhb}. The effective Lagrangian of the $H\gamma\gamma$, $HZZ$ and $H\gamma Z$ couplings in the SILH basis \cite{Giudice:2007ops} is described as the sum of the Lagrangian involving the CP-conserving and CP-violating dimension-six operators and the dimension-four SM Lagrangian as follows:

\begin{eqnarray}
\label{eq.1} 
{\cal L}={\cal L}_{\text{SM}}+\sum_{i}{\overline{c}_i}{\cal O}_i={\cal L}_{\text{SM}}+{\cal L}_{\text{CPC}}+{\cal L}_{\text{CPV}}
\end{eqnarray}

{\raggedright where ${\cal O}_i$ are the dimension-six operators and ${\overline{c}_i}$ are the Wilson coefficients. The effective Lagrangian with the CP-conserving and CP-violating operators corresponding to the second and third terms in Eq.~(\ref{eq.1}), respectively, are shown below \cite{Giudice:2007ops,Alloul:2014hws}}

\begin{eqnarray}
\label{eq.2} 
\begin{split}
{\cal L}_{\text{CPC}}=&\frac{\overline{c}_H}{2\upsilon^2}\partial^\mu (\Phi^\dagger \Phi)\partial_\mu(\Phi^\dagger \Phi)+\frac{\overline{c}_T}{2\upsilon^2}(\Phi^\dagger \overset\leftrightarrow{D}^\mu \Phi)(\Phi^\dagger \overset\leftrightarrow{D}_\mu \Phi) \\
&+\frac{ig\overline{c}_W}{m_W^2}(\Phi^\dagger T_{2k}\overset\leftrightarrow{D}^\mu\Phi)D^\nu W_{\mu\nu}^k+\frac{ig^\prime \overline{c}_B}{2m_W^2}(\Phi^\dagger \overset\leftrightarrow{D}^\mu\Phi) \partial^\nu B_{\mu\nu} \\
&+\frac{2ig\overline{c}_{HW}}{m_W^2}(D^\mu \Phi^\dagger T_{2k}D^\nu\Phi)W_{\mu\nu}^k+\frac{ig^\prime \overline{c}_{HB}}{m_W^2}(D^\mu \Phi^\dagger D^\nu\Phi)B_{\mu\nu} \\
&+\frac{g^{\prime2}\overline{c}_{\gamma}}{m_W^2} \Phi^\dagger \Phi B_{\mu\nu} B^{\mu\nu}
\end{split}
\end{eqnarray}

{\raggedright and}

\begin{eqnarray}
\label{eq.3} 
{\cal L}_{\text{CPV}}=\frac{ig\widetilde{c}_{HW}}{m_W^2}D^\mu \Phi^\dagger T_{2k} {D}^\nu\Phi \widetilde{W}_{\mu\nu}^k+\frac{ig^\prime \widetilde{c}_{HB}}{m_W^2}D^\mu \Phi^\dagger {D}^\nu\Phi \widetilde{B}_{\mu\nu}+\frac{g^{\prime 2} \widetilde{c}_{\gamma}}{m_W^2}\Phi^\dagger \Phi B_{\mu\nu} \widetilde{B}^{\mu\nu}
\end{eqnarray}

{\raggedright where $\upsilon$ is the vacuum expectation value of the Higgs field. $\widetilde{B}_{\mu\nu}=\frac{1}{2}\epsilon_{\mu\nu\rho\sigma}B^{\rho\sigma}$ and $\widetilde{W}_{\mu\nu}^k=\frac{1}{2}\epsilon_{\mu\nu\rho\sigma}W^{\rho\sigma k}$ are the dual field strength tensors defined by the field strength tensors ${B}_{\mu\nu}=\partial_\mu B_\nu - \partial_\nu B_\mu$ and ${W}_{\mu\nu}^k=\partial_\mu W_\nu^k - \partial_\nu W_\mu^k + g\epsilon_{ij}^k W_\mu^i W_\nu^j$ with gauge couplings $g^\prime$ and $g$. The $SU(2)_L$ generators are written as $T_{2k}=\sigma_k/2$, where $\sigma_k$ is the Pauli matrices. $D^\mu$ is covariant derivative operator with $\Phi^\dagger \overset\leftrightarrow{D}_\mu\Phi=\Phi^\dagger(D_\mu \Phi)-(D_\mu \Phi^\dagger)\Phi$, where $\Phi$ is the Higgs doublet in SM. In the unitarity gauge and mass basis, the Lagrangian with triple interaction terms with the Higgs sector for the $H\gamma\gamma$, $H\gamma Z$ and $HZZ$ couplings is given below \cite{Alloul:2014hws}:}

\begin{eqnarray}
\label{eq.4} 
\begin{split}
{\cal L}=&-\frac{1}{4}g_{h\gamma\gamma}F_{\mu\nu}F^{\mu\nu}h-\frac{1}{4}\widetilde{g}_{h\gamma\gamma}F_{\mu\nu}\widetilde{F}^{\mu\nu}h-\frac{1}{4}g_{hzz}^{(1)}Z_{\mu\nu}Z^{\mu\nu}h-g_{hzz}^{(2)}Z_{\nu}\partial_\mu Z^{\mu\nu}h \\
&+\frac{1}{2}g_{hzz}^{(3)}Z_{\mu}Z^{\mu}h-\frac{1}{4}\widetilde{g}_{hzz}Z_{\mu\nu}\widetilde{Z}^{\mu\nu}h-\frac{1}{2}g_{h\gamma z}^{(1)}Z_{\mu\nu}F^{\mu\nu}h-\frac{1}{2}\widetilde{g}_{h\gamma z}Z_{\mu\nu}\widetilde{F}^{\mu\nu}h-g_{h\gamma z}^{(2)}Z_{\nu}\partial_\mu F^{\mu\nu}h
\end{split}
\end{eqnarray}

{\raggedright where $h$, $F_{\mu\nu}$ and $Z_{\mu\nu}$ are the Higgs boson field, the field strength tensors of photon and $Z$-boson, respectively. $\widetilde{F}^{\mu\nu}=\frac{1}{2}\epsilon^{\mu\nu\rho\sigma}F_{\rho\sigma}$ and $\widetilde{Z}^{\mu\nu}=\frac{1}{2}\epsilon^{\mu\nu\rho\sigma}Z_{\rho\sigma}$ are the dual field strength tensors. The relationships among the Lagrangian parameters in the mass basis in Eq.~(\ref{eq.4}) and the gauge basis in Eqs.~(\ref{eq.2}-\ref{eq.3}) are shown below:}

\begin{eqnarray}
\label{eq.5} 
\widetilde{g}_{h\gamma\gamma}=-\frac{8g\widetilde{c}_\gamma s_W^2}{m_W}
\end{eqnarray}
\begin{eqnarray}
\label{eq.6} 
\widetilde{g}_{h\gamma z}=\frac{gs_W}{c_W m_W}\left[\widetilde{c}_{HW}-\widetilde{c}_{HB}+8\widetilde{c}_{\gamma}s_W^2\right]
\end{eqnarray}
\begin{eqnarray}
\label{eq.7} 
\widetilde{g}_{hzz}=\frac{2g}{c_W^2 m_W}\left[\widetilde{c}_{HB}s_W^2-4\widetilde{c}_{\gamma}s_W^4+c_W^2\widetilde{c}_{HW}\right]
\end{eqnarray}

{\raggedright for CP-violating couplings and}

\begin{eqnarray}
\label{eq.8} 
g_{h\gamma\gamma}=a_H-\frac{8g\overline{c}_\gamma s_W^2}{m_W}
\end{eqnarray}
\begin{eqnarray}
\label{eq.9} 
g_{h\gamma z}^{(1)}=\frac{gs_W}{c_Wm_W}\left[\overline{c}_{HW}-\overline{c}_{HB}+8\overline{c}_{\gamma}s_W^2\right]
\end{eqnarray}
\begin{eqnarray}
\label{eq.10} 
g_{h\gamma z}^{(2)}=\frac{gs_W}{c_Wm_W}\left[\overline{c}_{HW}-\overline{c}_{HB}-\overline{c}_{B}+\overline{c}_{W}\right]
\end{eqnarray}
\begin{eqnarray}
\label{eq.11} 
g_{hzz}^{(1)}=\frac{2g}{c_W^2 m_W}\left[\overline{c}_{HB}s_W^2-4\overline{c}_{\gamma}s_W^4+c_W^2 \overline{c}_{HW}\right]
\end{eqnarray}
\begin{eqnarray}
\label{eq.12} 
g_{hzz}^{(2)}=\frac{g}{c_W^2 m_W}\left[(\overline{c}_{HW}+\overline{c}_{W})c_W^2+(\overline{c}_B+\overline{c}_{HB})s_W^2\right]
\end{eqnarray}
\begin{eqnarray}
\label{eq.13} 
g_{hzz}^{(3)}=\frac{gm_W}{c_W^2}\left[1-\frac{1}{2}\overline{c}_{H}-2\overline{c}_{T}+8\overline{c}_\gamma \frac{s_W^4}{c_W^2}\right]
\end{eqnarray}

{\raggedright for CP-conserving couplings. Here, $a_H$ represents the SM contribution at the Higgs boson to two photons vertex, $c_W=\text{cos}\theta_W$, $s_W=\text{sin}\theta_W$ and $\theta_W$ is the weak mixing angle.}

As seen in Eqs.~(\ref{eq.5}-\ref{eq.13}), the Higgs-electroweak gauge boson couplings in the effective Lagrangian in the mass basis for the anomalous $H\gamma\gamma$, $HZZ$ and $H\gamma Z$ couplings are sensitive to the ten Wilson coefficients related to the Higgs-gauge boson couplings. These coefficients $\overline{c}_\gamma$, $\overline{c}_B$, $\overline{c}_W$, $\overline{c}_{HB}$, $\overline{c}_{HW}$, $\overline{c}_T$, $\overline{c}_H$ associate with CP-conserving couplings and coefficients $\widetilde{c}_\gamma$, $\widetilde{c}_{HB}$, $\widetilde{c}_{HW}$ associate with CP-violating couplings.

We have performed analyses of the effects of dimension-six operators on Higgs-gauge boson couplings in {\sc MadGraph5}$\_$aMC@NLO \cite{Alwall:2014cvc} based on Monte Carlo simulations using FeynRules \cite{Alloul:2014tfc}. In the Higgs effective Lagrangian model file of the FeynRules, the SILH Lagrangian is included along with the Wilson coefficients for Higgs boson interactions.

We investigate the potential of anomalous Higgs-gauge boson couplings on the process $\mu^- \mu^+ \rightarrow Z\gamma\gamma \rightarrow \ell \ell \gamma \gamma$ at the muon collider and the sensitivity study of coefficients $\overline{c}_\gamma$, $\widetilde{c}_\gamma$, $\overline{c}_{HB}$, $\widetilde{c}_{HB}$, $\overline{c}_{HW}$ and $\widetilde{c}_{HW}$ in anomalous $H\gamma\gamma$, $HZZ$ and $H\gamma Z$ vertices. The stage of the muon collider with a center-of-mass energy of 10 TeV and an integrated luminosity of 10 ab$^{-1}$ is considered \cite{Accettura:2023uhz}. By combining high-energy collisions in a clean experimental environment, the multi-TeV muon collider creates unique potential for Higgs studies by providing both precise measurements and high-energy reach in a single machine. Therefore, a 10 TeV lepton collider that goes beyond the precision-energy dilemma is the ideal machine. Since electron-based colliders using current technology are unable to reach this energy scale, we are turning to muon colliders to probe this scale with leptons \cite{Black:2024edf}.

\begin{figure}[H]
\centering
\includegraphics[scale=0.63]{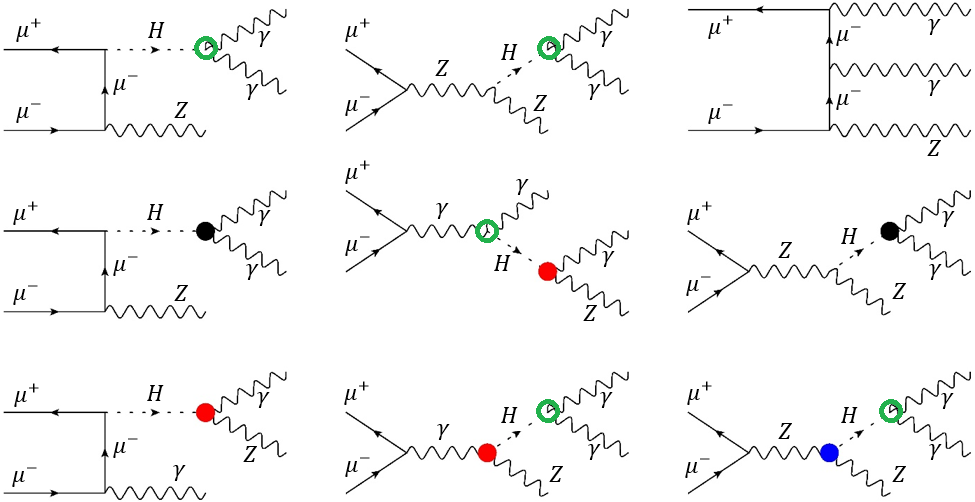}
\caption{The Feynman diagrams of the process $\mu^- \mu^+ \rightarrow Z\gamma\gamma$. 
\label{fig1}}
\end{figure}

The Feynman diagrams of the process $\mu^- \mu^+ \rightarrow Z\gamma\gamma$ are shown in Fig.~\ref{fig1}. The three diagrams in the top row are pure SM background processes, while the other six diagrams present signal processes, including new physics contributions from the anomalous $H\gamma\gamma$ (black dot), $H\gamma Z$ (red dot) and $HZZ$ (blue dot) vertices.
The $H\gamma\gamma$ vertex (green circle) does not exist at the tree-level in the SM, but is present at the loop-level and contributes to the term $a_H$ in Eq.~\ref{eq.8}.

\begin{figure}[H]
\centering
\includegraphics[scale=0.7]{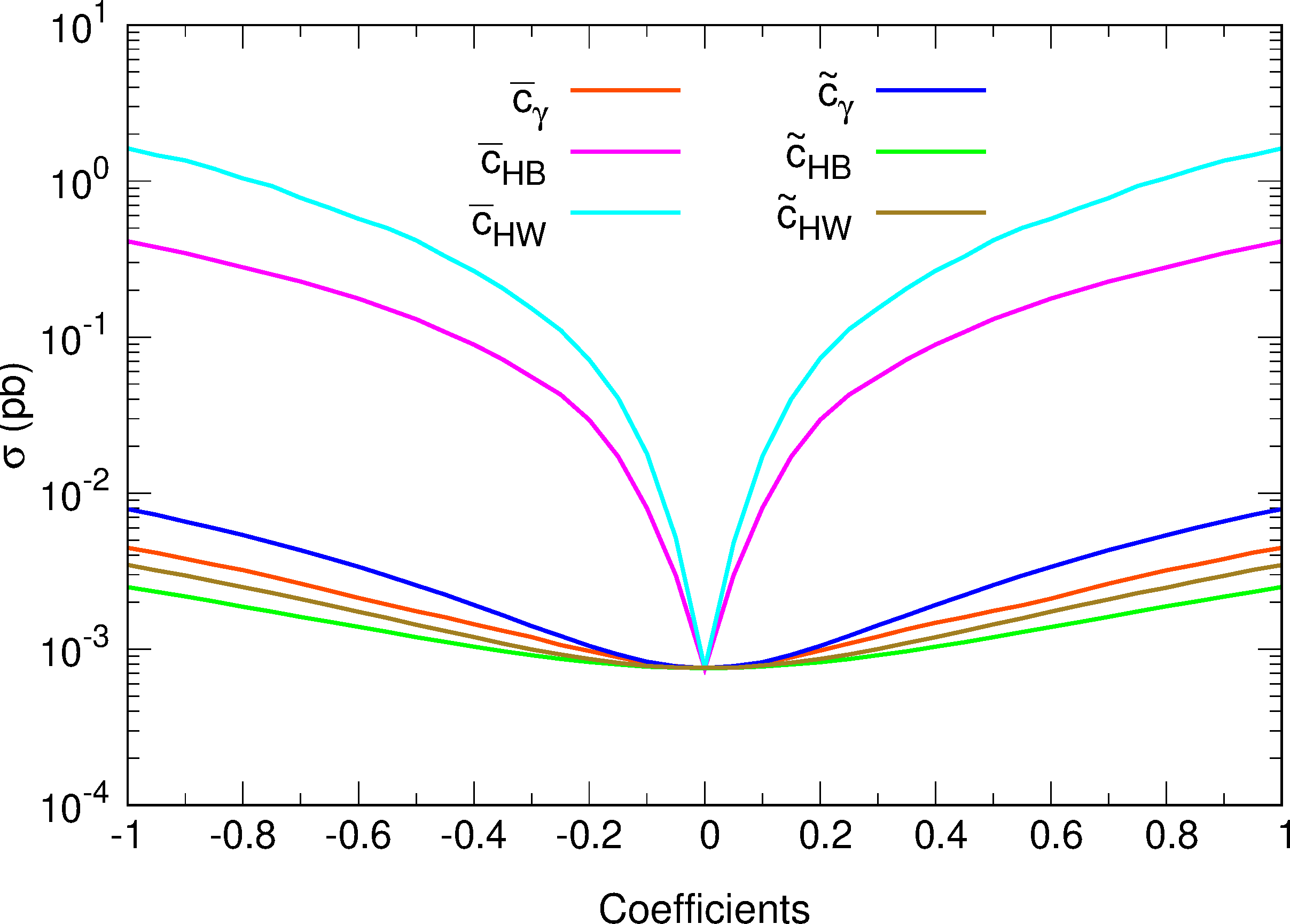}
\caption{The total cross-section of the process $\mu^- \mu^+ \rightarrow Z\gamma\gamma$ as a function of coefficients $\overline{c}_\gamma$, $\widetilde{c}_\gamma$, $\overline{c}_{HB}$, $\widetilde{c}_{HB}$, $\overline{c}_{HW}$ and $\widetilde{c}_{HW}$ at the muon collider.}
\label{fig2}
\end{figure}

The total cross-sections of the process $\mu^- \mu^+ \rightarrow Z\gamma\gamma$ as a function of coefficients $\overline{c}_\gamma$, $\widetilde{c}_\gamma$, $\overline{c}_{HB}$, $\widetilde{c}_{HB}$, $\overline{c}_{HW}$ and $\widetilde{c}_{HW}$ at the muon collider are given in Fig.~\ref{fig2}. The transverse momentum $p_T^{\gamma} > 10$ GeV and the pseudo-rapidity $|\eta^{\gamma}| < 2.5$ for the photons on the final state particles are applied to determine the total cross-sections in Fig.~\ref{fig2}. The total cross-sections in Fig.~\ref{fig2} were created by calculating a function of the coefficient in question with Madgraph when all other coefficients were zero. The SM cross-section including the contributions from the top three Feynman diagrams in Fig.~\ref{fig1} corresponds to the point $\overline{c}_\gamma=\overline{c}_{HB}=\overline{c}_{HW}=\widetilde{c}_\gamma=\widetilde{c}_{HB}=\widetilde{c}_{HW}=0$ in Fig.~\ref{fig2}.

It shows that the cross-sections as a function of the coefficients $\overline{c}_{HB}$ and $\overline{c}_{HW}$ are larger than those of the coefficients $\overline{c}_\gamma$, $\widetilde{c}_\gamma$, $\widetilde{c}_{HB}$ and $\widetilde{c}_{HW}$. That is, the process of $Z\gamma\gamma$ production is not sensitive to changes in the coefficients $\overline{c}_\gamma$, $\widetilde{c}_\gamma$, $\widetilde{c}_{HB}$ and $\widetilde{c}_{HW}$, but only to $\overline{c}_{HB}$ and $\overline{c}_{HW}$. Therefore, although we will present the effects of these six coefficients at the muon collider in the following sections, for the purpose of this paper we will focus on the coefficients $\overline{c}_{HB}$ and $\overline{c}_{HW}$.

\section{Analysis of Signal and Background Events} \label{Sec3}

We describe the cut-based analysis and the detector simulation to determine the limits on anomalous Higgs-gauge boson couplings for the $H\gamma\gamma$, $HZZ$ and $H\gamma Z$ vertices at the muon collider. The process $\mu^- \mu^+ \rightarrow Z\gamma\gamma \rightarrow \ell \ell \gamma \gamma$ with nonzero coefficients $\overline{c}_\gamma$, $\widetilde{c}_\gamma$, $\overline{c}_{HB}$, $\widetilde{c}_{HB}$, $\overline{c}_{HW}$, $\widetilde{c}_{HW}$ is considered signal process that comprises the SM contribution as well as interference between effective couplings and SM contributions. The following five relevant backgrounds are considered. (i) The SM contribution from the same final state of the signal process $\mu^- \mu^+ \rightarrow Z\gamma\gamma \rightarrow \ell \ell \gamma \gamma$, where the $Z$-boson decays into a pair of charged leptons with the same flavor and opposite charges ($\ell = e, \mu$), is considered the main background process. (ii) The process $\mu^- \mu^+ \rightarrow \tau \tau \gamma \gamma$ is considered a background process. (iii) Owing to the similarity in shape of the electromagnetic clusters between photons and electrons, a significant fraction of electrons can be mistakenly identified as photons. The background process $\mu^- \mu^+ \rightarrow ZZ \rightarrow \ell \ell \ell \ell$, in which electrons are misidentified as photons, is considered. (iv) Another background contribution comes from the $t\bar{t}\gamma\gamma$ events, in which the top quarks decay leptonically in the $\mu^- \mu^+ \rightarrow t\bar{t}\gamma\gamma$ process. (v) The background process containing jets that are misidentified as photons is located in the $\ell \ell \gamma \gamma$ signal region. In process $\mu^- \mu^+ \rightarrow Zjj \rightarrow \ell \ell jj$, both photon candidates are misidentified jets.

In {\sc MadGraph5}$\_$aMC@NLO, 500k events at $\sqrt{s}=10$ TeV are generated for different values of coefficients $\overline{c}_\gamma$, $\widetilde{c}_\gamma$, $\overline{c}_{HB}$, $\widetilde{c}_{HB}$, $\overline{c}_{HW}$, $\widetilde{c}_{HW}$ of the signal process and for the background processes in all analyses of this study. These events, which involve initial and final parton shower, fragmentation and decay, are processed through the Pythia 8.3 \cite{Bierlich:2022uzx}. The Delphes 3.5.0 package \cite{Favereau:2014wfb}, which allows the simulation of many multipurpose detectors, is used to simulate the realistic detector effects of the muon collider. At electron energy $E \geq 400$ GeV, the electron efficiencies are 68\% for $1.95 <|\eta|< 2.5$, 96\% for $1.22 <|\eta|\leq 1.95$, 82\% for $1.1 <|\eta|\leq 1.22$, 96\% for $0.91 <|\eta|\leq 1.1$, 98\% for $0.69 <|\eta|\leq 0.91$ and 98\% for $|\eta|\leq 0.69$. At muon energy $E \geq 2.5$ GeV, the muon efficiency is 99.9\%. At photon energy $E \geq 2$ GeV, the photon efficiencies are 94\% for $|\eta|< 0.7$ and 90\% for $0.7 \leq |\eta|\leq 2.5$. In addition, many performance parameters regarding the detector simulation of the muon collider at Delphes, such as tracking efficiency, resolution, electron/muon/photon efficiency and tagging, are discussed in Ref.~\cite{Selvaggi:2020yhe}. Furthermore, all events are analyzed with ROOT 6 \cite{Brun:1997gqa}.

\begin{figure}[H]
\centering
\begin{subfigure}{0.47\linewidth}
\includegraphics[width=\linewidth]{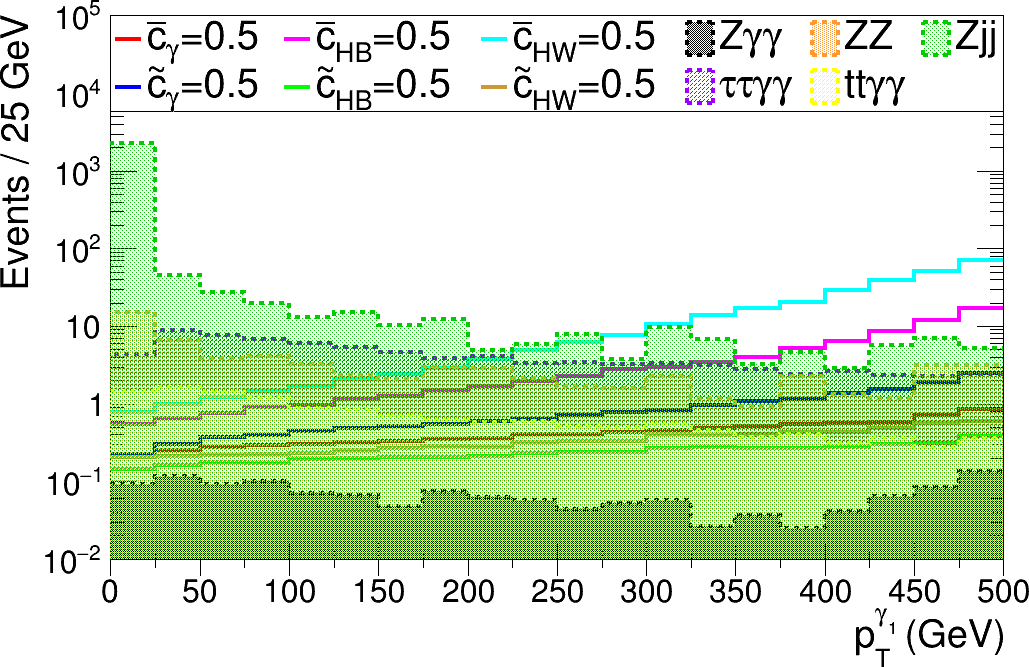}
\caption{}
\label{fig3:a}
\end{subfigure}\hfill
\begin{subfigure}{0.47\linewidth}
\includegraphics[width=\linewidth]{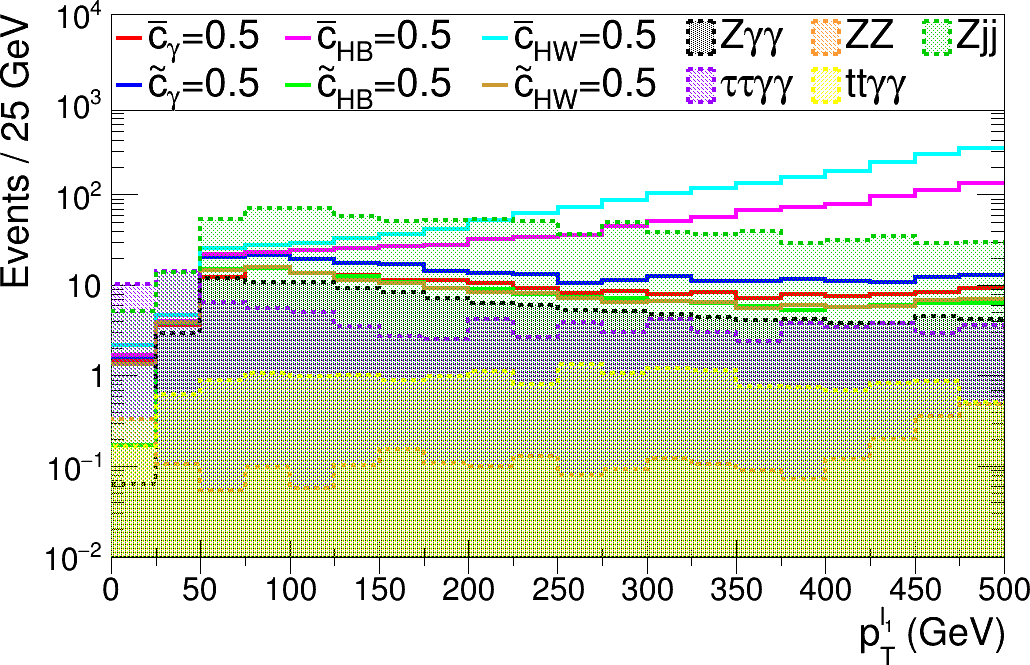}
\caption{}
\label{fig3:b}
\end{subfigure}\hfill

\begin{subfigure}{0.47\linewidth}
\includegraphics[width=\linewidth]{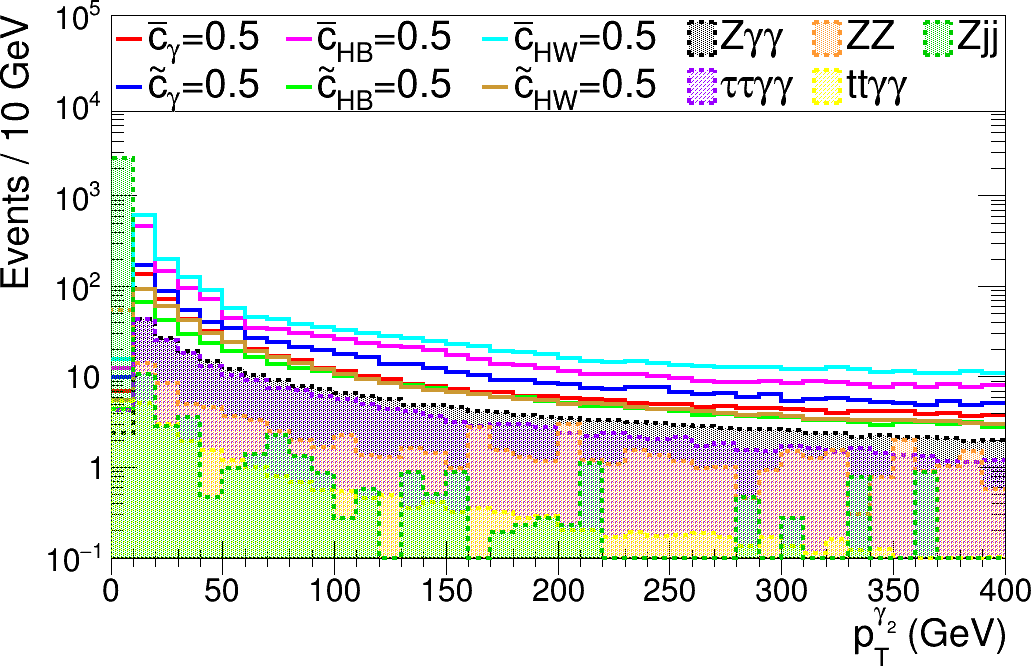}
\caption{}
\label{fig3:c}
\end{subfigure}\hfill
\begin{subfigure}{0.47\linewidth}
\includegraphics[width=\linewidth]{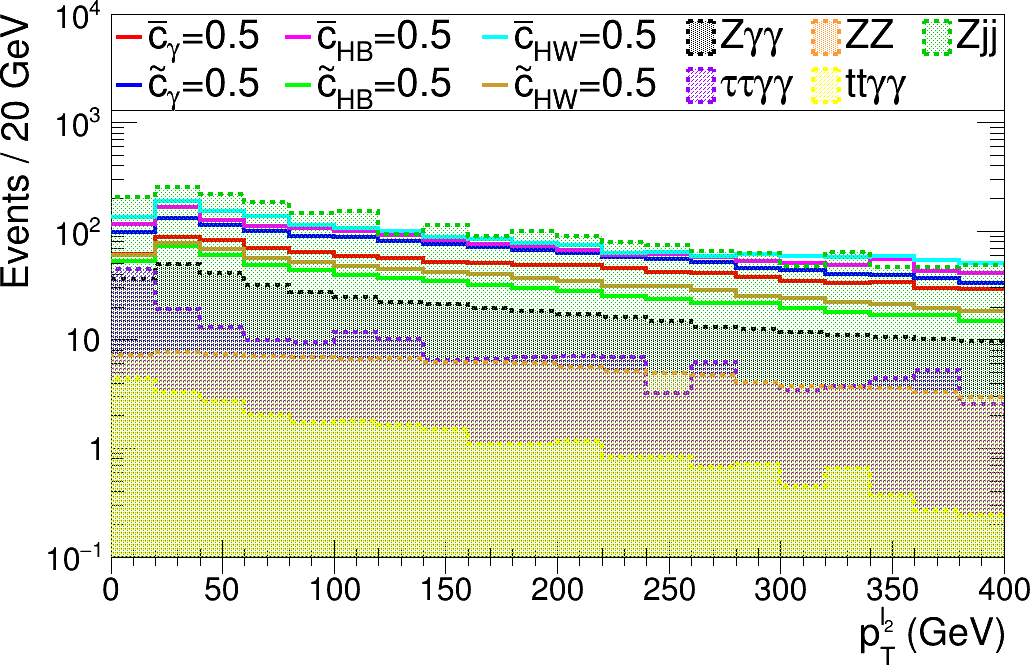}
\caption{}
\label{fig3:d}
\end{subfigure}\hfill

\caption{The distributions of transverse momentum of the leading photon (a), the leading charged lepton (b), the sub-leading photon (c) and the sub-leading charged lepton (d) for signal and relevant background
processes at the muon collider.}
\label{fig3}
\end{figure}

A first preselection is applied on events, requiring at least two photons in the $Z\gamma \gamma$ production and at least two leptons with the same flavour and opposite-charge in the decay of the $Z$-boson and this preselection is labeled with Cut-0. According to their transverse momentum, the leading and sub-leading charged leptons ($\ell_1$ and $\ell_2$) and photons ($\gamma_1$ and $\gamma_2$) are arranged as follows: $p_T^{\ell_1} > p_T^{\ell_2}$ and $p_T^{\gamma_1} > p_T^{\gamma_2}$, respectively. Fig.~\ref{fig3} shows the distributions of the transverse momentum of the leading photon and the leading charged lepton for the signal and the relevant background processes. It can be seen that the signal can be distinguished from the backgrounds by $p_T^{\gamma_1} > 50$ GeV and $p_T^{\ell_1} > 50$ GeV. The transverse momentum and pseudo-rapidity distributions of the leading and sub-leading photons for the events are required to be $p_T^{\gamma_1} > 50$ GeV, $p_T^{\gamma_2} > 10$ GeV and $|\eta^{\gamma_{1,2}}|\leq 2.5$, respectively, and are labeled Cut-1. However, the distributions of transverse momentum and pseudo-rapidity of the leading and sub-leading charged leptons for the events are required to be $p_T^{\ell_1} > 50$ GeV, $p_T^{\ell_2} > 20$ GeV and $|\eta^{\ell_{1,2}}|\leq 2.5$, respectively, and are labeled Cut-2.

\begin{figure}[H]
\centering
\begin{subfigure}{0.48\linewidth}
\includegraphics[width=\linewidth]{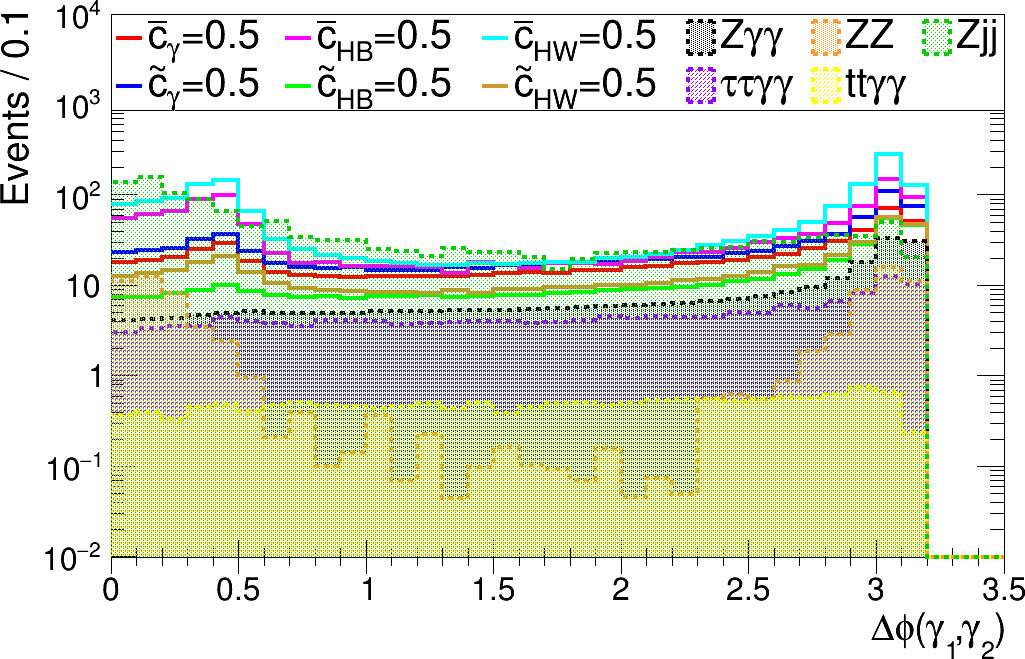}
\caption{}
\label{fig4:a}
\end{subfigure}\hfill
\begin{subfigure}{0.48\linewidth}
\includegraphics[width=\linewidth]{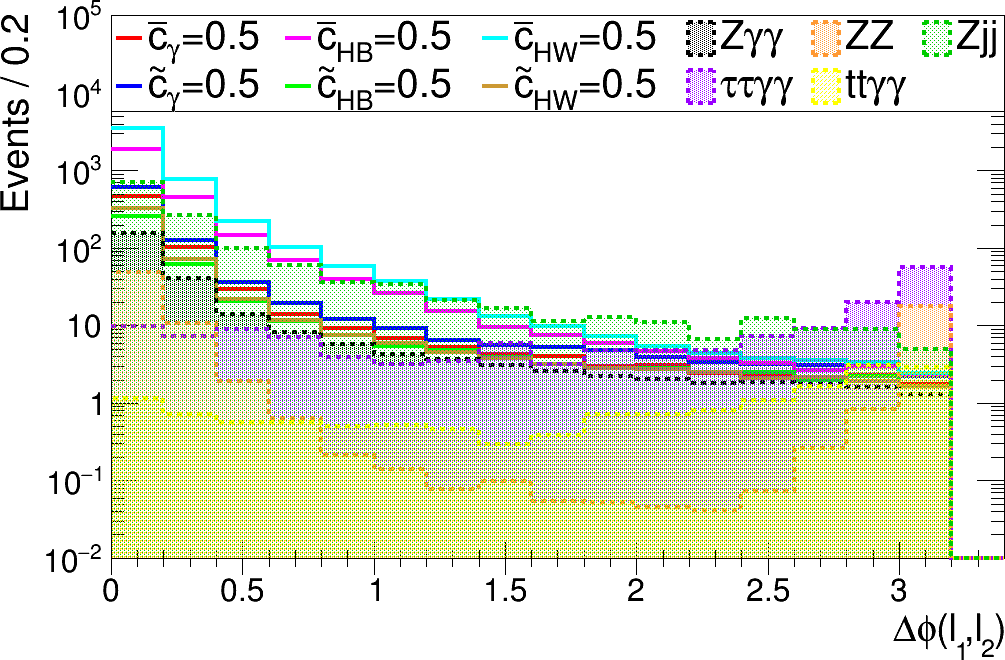}
\caption{}
\label{fig4:b}
\end{subfigure}\hfill

\caption{The distributions of azimuthal angle between the two photons (left) and two charged leptons (right) for signal and relevant background processes at the muon collider.}
\label{fig4}
\end{figure}

\begin{figure}[H]
\centering
\begin{subfigure}{0.48\linewidth}
\includegraphics[width=\linewidth]{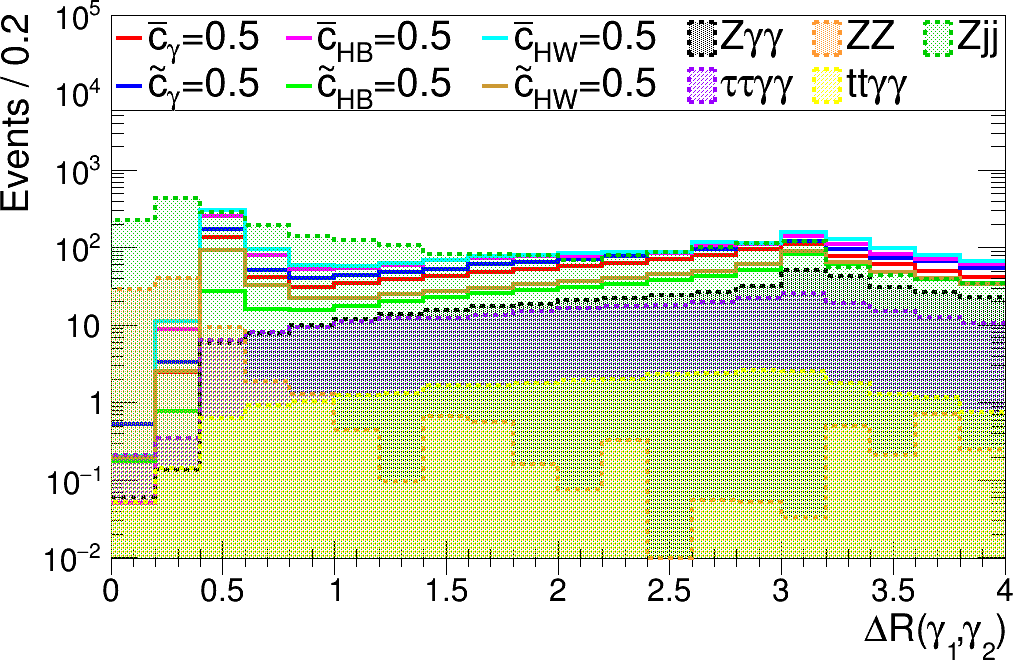}
\caption{}
\label{fig5:a}
\end{subfigure}\hfill
\begin{subfigure}{0.48\linewidth}
\includegraphics[width=\linewidth]{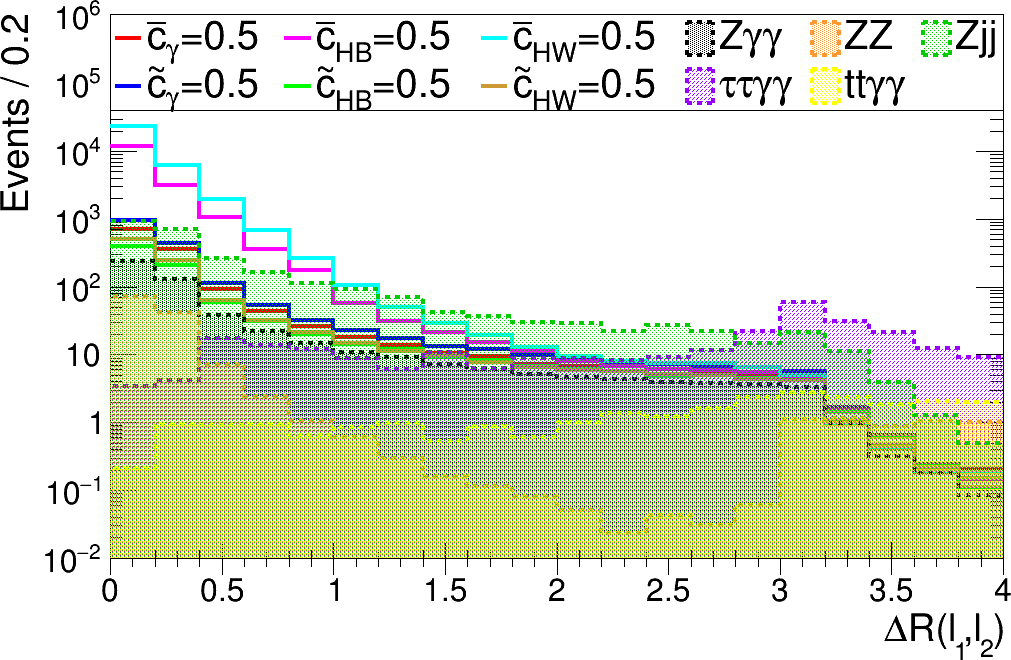}
\caption{}
\label{fig5:b}
\end{subfigure}\hfill
\begin{subfigure}{0.48\linewidth}
\includegraphics[width=\linewidth]{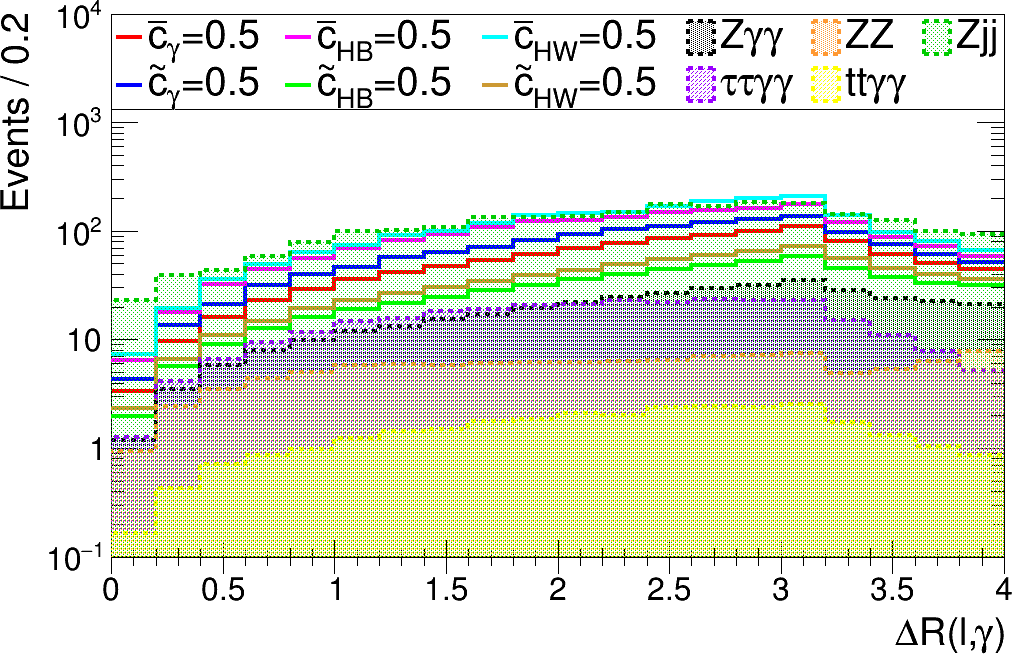}
\caption{}
\label{fig5:c}
\end{subfigure}\hfill

\caption{The distributions of the angular distance $\Delta R_{\gamma \gamma}$ (a), $\Delta R_{\ell \ell}$ (b) and $\Delta R_{\ell \gamma}$ (c) for signal and relevant background processes at the muon collider.}
\label{fig5}
\end{figure}

A clear difference between signal and background is observed in the distribution of the azimuthal angle between the leading and the sub-leading photon, the azimuthal angle between the two charged leptons in Fig.~\ref{fig4}. The angular distance between two particles at the detector in the cylindrical coordinate system in the $\eta – \phi$ space is given by $\Delta R=\sqrt{|\Delta\eta|^2+|\Delta\phi|^2}$, where $\Delta\eta$ and $\Delta\phi$ are the pseudo-rapidity and the azimuthal angle differences, respectively. The distributions of the angular distance between the two photons, between the two charged leptons, and between the leading charged lepton and the leading photon are shown in Fig.~\ref{fig5}. For photons and charged leptons to be well separated in phase space leading to their identification as distinct objects in the detector, kinematic cuts are needed such as $\Delta \phi_{\gamma \gamma} < 0.8$ and $\Delta \phi_{\ell \ell} < 1.8$ at Cut-3 and $\Delta R_{\gamma \gamma} > 0.4$, $\Delta R_{\ell \ell} < 1.4$ and $\Delta R_{\ell \gamma} > 0.4$ at Cut-4.

\begin{figure}[H]
\centering
\begin{subfigure}{0.48\linewidth}
\includegraphics[width=\linewidth]{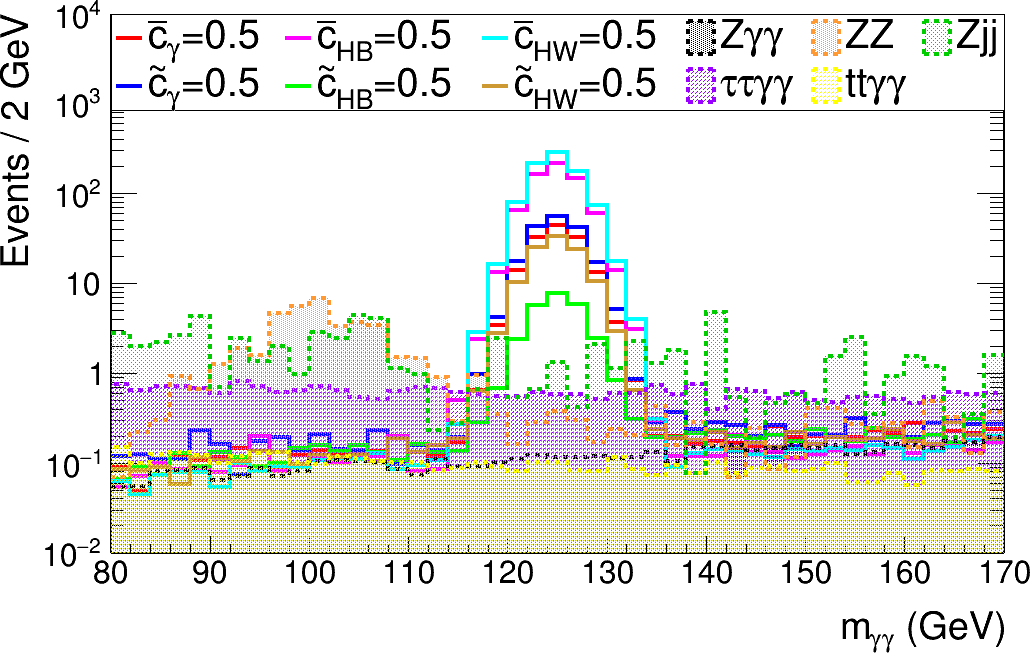}
\caption{}
\label{fig6:a}
\end{subfigure}\hfill
\begin{subfigure}{0.48\linewidth}
\includegraphics[width=\linewidth]{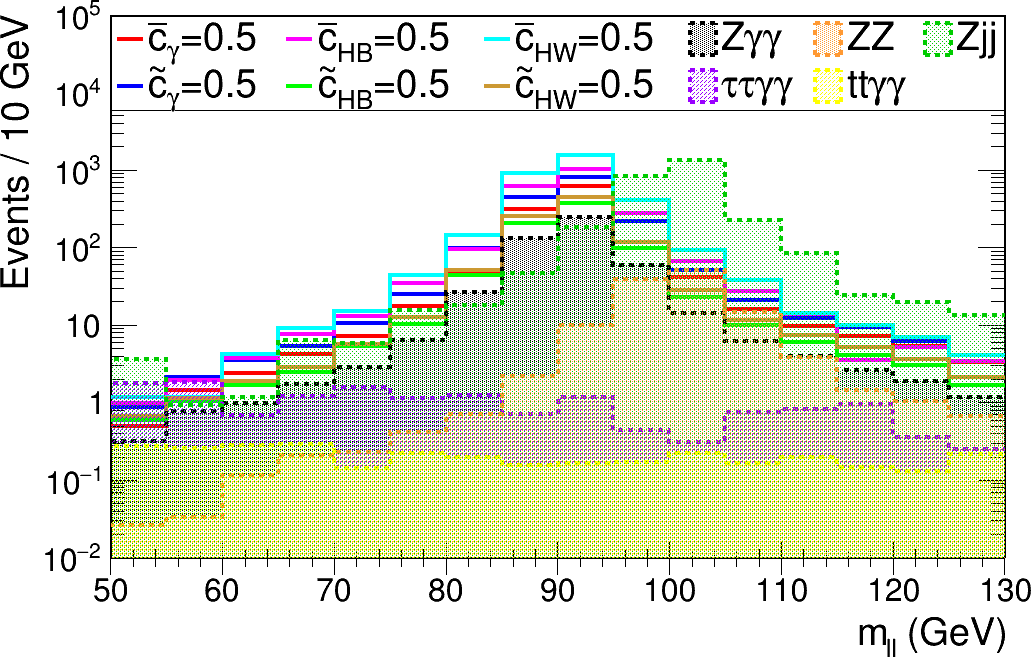}
\caption{}
\label{fig6:b}
\end{subfigure}\hfill

\caption{The invariant mass distributions of photon pair (left) and charged lepton pair (right) for signal and relevant background processes at the muon collider.}
\label{fig6}
\end{figure}

In order to separate the signal from the backgrounds in the invariant mass distribution of the photon pair and the lepton pair in Fig.~\ref{fig6}, we set Cut-5 as $110$ GeV$ < m_{\gamma \gamma} < 140$ GeV and $|m_{\ell\ell}-m_Z| < 20$ GeV, where $m_Z=91.2$ GeV. The mass of the Higgs boson is approximately 125 GeV. The decay of the Higgs into two photons is a rare but distinguishable process. The invariant mass of the two photons, the products of this decay, becomes equal to the mass of the Higgs. Fig.~\ref{fig6:a} shows a peak in the $m_{\gamma \gamma}$ distribution around 125 GeV. This peak indicates that the Higgs boson is produced in the signal process and is critical for the experimental detection of the existence of the Higgs boson. The reason why we choose a wide range such as $110$ GeV$ < m_{\gamma \gamma} < 140$ GeV instead of applying a narrow cut is to increase the contribution of $H\gamma\gamma$ vertex without completely losing the contribution of $HZ\gamma$ vertex. The mass of the $Z$-boson is approximately 91.2 GeV. Since the $Z$-boson decays to two charged leptons during the signal process, the $m_{\ell\ell}$ distribution in Fig.~\ref{fig6:b} shows the most events around $m_{\ell\ell}\approx 91.2$ GeV. This is called the resonance peak. This peak indicates that a real $Z$-boson was produced in the experiment and that it decays to a lepton pair. Lepton pairs from background events can have different invariant masses, but events from the $Z$-boson are heavily concentrated at 91.2 GeV. In summary, Table~\ref{tab1} lists all the kinematic cuts of the analysis.

\begin{table}[H]
\centering
\caption{Applied kinematic cuts in the analysis at the muon collider.}
\label{tab1}
\begin{tabular}{p{2.5cm}p{10.7cm}}
\hline \hline
Cut flow & Definitions \\
\hline
Cut-0 & $N_{\gamma} \geq 2$, $N_{\ell} \geq 2$ (least two same flavor opposite-charge leptons)\\ 
Cut-1 & $p_T^{\gamma_1} > 50$ GeV, $p_T^{\gamma_2} > 10$ GeV, $|\eta^{\gamma_{1,2}}| < 2.5$\\
Cut-2 & $p_T^{\ell_1} > 50$ GeV, $p_T^{\ell_2} > 20$ GeV, $|\eta^{\ell_{1,2}}| < 2.5$\\ 
Cut-3 & $\Delta \phi_{\gamma \gamma} < 0.8$, $\Delta \phi_{\ell \ell} < 1.8$\\
Cut-4 & $\Delta R_{\gamma \gamma} > 0.4$, $\Delta R_{\ell \ell} < 1.4$, $\Delta R_{\ell \gamma} > 0.4$\\
Cut-5 & $110$ GeV$ < m_{\gamma \gamma} < 140$ GeV, $|m_{\ell\ell}-m_Z| < 20$ GeV\\  \hline \hline
\end{tabular}
\end{table}

The cumulative event numbers of the signal (in turn, each coefficient is set equal to 0.5, while all other coefficients are set to zero) and relevant backgrounds ($Z\gamma\gamma$, $\tau\tau\gamma\gamma$, $ZZ$, $t\bar{t}\gamma\gamma$, $Zjj$) after each cut used in the analysis are given in Table~\ref{tab2}. In this table, the cumulative event numbers are calculated by multiplying the cross-sections generated in MadGraph5 passing through Pythia and using Delphes with the integrated luminosity, where the integrated luminosity of the muon collider is ${\cal L}_{\text{int}}=10$ ab$^{-1}$. The signal to total background ratio ($S/B$) after each cut is presented in Table~\ref{tab2} to examine the efficiency of kinematic cuts in suppressing backgrounds. The impact of Cut-5 plays a significant role in suppressing relevant backgrounds by increasing the signal to total background ratio in cut flow by approximately 20 times.

\begin{table}[H]
\centering
\caption{The cumulative number of events for signal and relevant background processes after applied kinematic cuts at muon collider.}
\label{tab2}
\begin{tabular}{p{1.9cm}p{1.1cm}p{1.1cm}p{1.1cm}p{1.1cm}p{1.1cm}p{1.0cm}p{1.1cm}p{1.0cm}p{1.1cm}p{1cm}p{1.1cm}p{1.0cm}}
\hline \hline
& \multicolumn{2}{c}{Cut-0} & \multicolumn{2}{c}{Cut-1} & \multicolumn{2}{c}{Cut-2} & \multicolumn{2}{c}{Cut-3} & \multicolumn{2}{c}{Cut-4} & \multicolumn{2}{c}{Cut-5}\\ [-0.5ex] \hline
Signal & Event & $S/B$ & Event & $S/B$ & Event & $S/B$ & Event & $S/B$ & Event & $S/B$ & Event & $S/B$\\ [-0.5ex] \hline
$\overline{c}_{\gamma}=0.5$ & 123 & 0.043 & 120 & 0.472 & 111 & 0.539 & 77 & 0.696 & 66 & 1.183 & 23 & 17.35\\ 
$\widetilde{c}_{\gamma}=0.5$ & 174 & 0.061 & 170 & 0.669 & 152 & 0.737 & 106 & 0.959 & 92 & 1.649 & 33 & 24.91\\ 
$\overline{c}_{HB}=0.5$ & 10867 & 3.813 & 10623 & 41.82 & 8900 & 43.20 & 8493 & 76.83 & 7530 & 134.9 & 6461 & 4876\\  
$\widetilde{c}_{HB}=0.5$ & 98 & 0.034 & 96 & 0.377 & 84 & 0.407 & 52 & 0.470 & 44 & 0.788 & 4.37 & 3.298\\  
$\overline{c}_{HW}=0.5$ & 25086 & 8.802 & 24722 & 97.33 & 21780 & 105.7 & 20847 & 188.6 & 18695 & 335.1 & 15605 & 11777\\  
$\widetilde{c}_{HW}=0.5$ & 111 & 0.039 & 109 & 0.429 & 97 & 0.471 & 66 & 0.597 & 56 & 1.003 & 18 & 13.58\\  \hline
Backgrounds & & & & & & & & & & & &\\ [-0.5ex] \hline
$Z\gamma\gamma$ & 70 && 69 && 60 && 35 && 30 && 0.27 &\\  
$\tau\tau\gamma\gamma$ & 133 && 119 && 86 && 36 && 12 && 0.06\\
$ZZ$ & 59 && 26 && 25 && 20 && 4.87 && 0.14\\
$t\bar{t}\gamma\gamma$ & 23 && 17 && 14 && 5.54 && 1.20 && 0.005\\
$Zjj$ & 2565 && 23 && 21 && 14 && 7.71 && 0.85\\
\hline \hline
\end{tabular}
\end{table}

\section{Sensitivities on the Higgs-gauge boson couplings} \label{Sec4}

The $\chi^2$ test is used to investigate the sensitivity of Higgs-gauge boson couplings in the $\mu^- \mu^+ \rightarrow Z\gamma\gamma \rightarrow \ell \ell \gamma \gamma$ process. The $\chi^2$ distribution, where the critical value of $\chi^2$ corresponding to one degree of freedom is equal to 3.84, is defined as follows in order to achieve limits at the 95\% Confidence Level (C.L.):

\begin{eqnarray}
\label{eq.14}
\chi^{2}=\sum_{i}^{n_{bins}} (\frac{N_{i}^{TOT}-N_{i}^{B}}{N_{i}^{B}\Delta_{i}})^{2}
\end{eqnarray}

\begin{figure}[H]
\centering
\begin{subfigure}{0.48\linewidth}
\includegraphics[width=\linewidth]{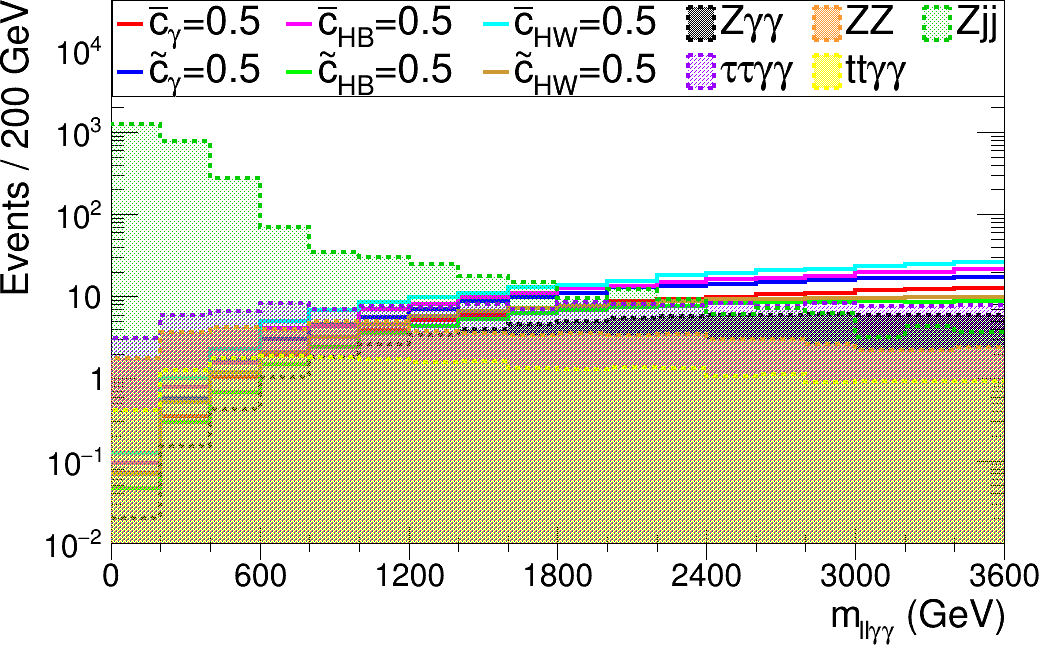}
\caption{}
\label{fig7:a}
\end{subfigure}\hfill
\begin{subfigure}{0.48\linewidth}
\includegraphics[width=\linewidth]{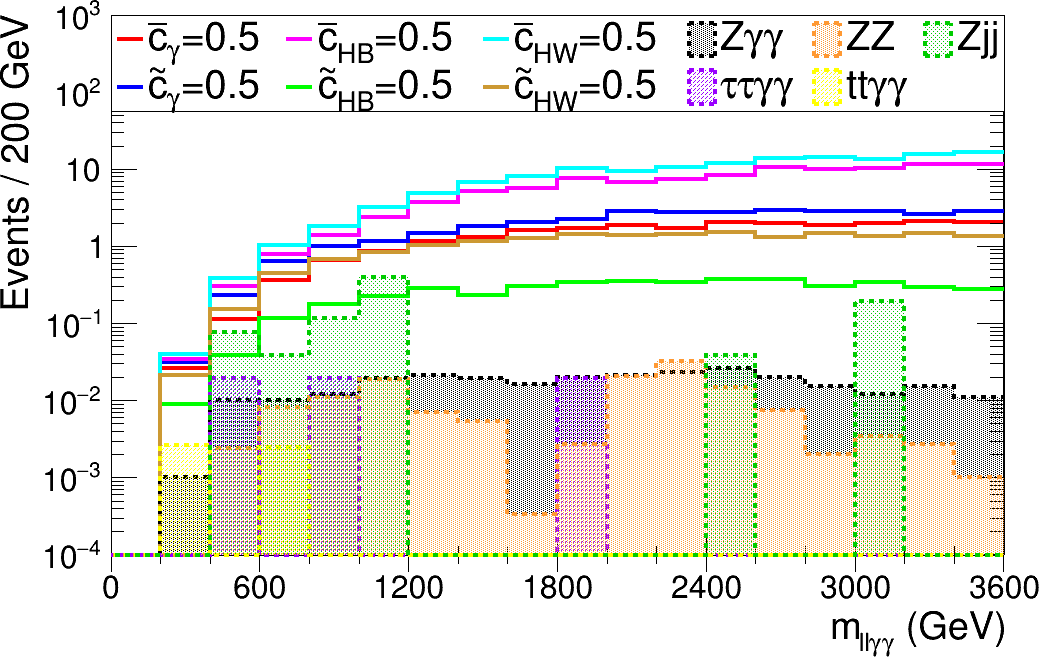}
\caption{}
\label{fig7:b}
\end{subfigure}\hfill

\caption{The invariant mass $m_{\ell \ell\gamma\gamma}$ distribution of $\ell^+ \ell^- \gamma\gamma$ final state after Cut-0 (left) and after Cut-5 (right) for signal and relevant background processes at the muon collider.}
\label{fig7}
\end{figure}

{\raggedright where $N_{i}^{B}$ is the number of events of relevant backgrounds and $N_{i}^{TOT}$, as a function of Wilson coefficients, is the total number of events including contributions of effective couplings in ith bin of the invariant mass $m_{\ell \ell\gamma\gamma}$ distribution of $\ell^+ \ell^- \gamma\gamma$ system. $\Delta_{i}=\sqrt{\delta_{sys}^2+\delta_{st}^2}$ includes systematic uncertainty $(\delta_{sys})$ and statistical uncertainty in each bin $(\delta_{st}=1/\sqrt{N_i^B})$ \cite{Spor:2022oer}. Fig.~\ref{fig7:a} and Fig.~\ref{fig7:b} show the invariant mass $m_{\ell \ell\gamma\gamma}$ distribution of $\ell^+ \ell^- \gamma\gamma$ system after Cut-0 (left) and after all applied cuts (right) for the signal and relevant background processes at the muon collider.}

Although the center-of-mass energy is 10 TeV, the process $\mu^- \mu^+ \rightarrow Z\gamma\gamma \rightarrow \ell \ell \gamma \gamma$ is not a direct 10 TeV resonance production. The $Z$-boson and photons produced in this process do not carry all the center-of-mass energy; most of the event is carried by particles and radiation traveling in very forward directions. In high-energy collisions, some particles and radiation are scattered along the beamline with small angles and cannot be detected by forward detectors if they are outside the acceptance range, thus increasing energy losses. In particular, a significant amount of energy may not be captured by the detector due to the initial-state radiation (ISR) and final-state radiation (FSR). Although muons radiate less than electrons, these effects cannot be ignored at very high energies, such as 10 TeV. In short, the reason why the observed $m_{\ell \ell \gamma \gamma}$ distribution does not peak at 10 TeV is that due to the nature of the process, the total energy of the observed final-state particles is much lower than the center-of-mass energy in events and the unobserved radiation plays an important role. 

The median expected significance $Z$ for the signal discovery is computed as \cite{Cowan:2011wer,Kumar:2015tgv,Senol:2024umn}:

\begin{eqnarray}
\label{eq.15}
Z=\sqrt{2[(S+B)\text{ln}(\frac{(S+B)(1+\delta^2 B)}{B+\delta^2 B(S+B)})-\frac{1}{\delta^2}\text{ln}(1+\frac{\delta^2 S}{1+\delta^2 B})]}
\end{eqnarray}

{\raggedright where $\delta$ is the systematic uncertainty. $S$ and $B$ are the numbers of events obtained by the invariant mass $m_{\ell \ell\gamma\gamma}$ distribution of $\ell^+ \ell^- \gamma\gamma$ system for the signal and the sum of the relevant backgrounds, respectively. In the case where the background expectation is known exactly, with the limit of $\delta \rightarrow 0$, this expression simplifies to:}

\begin{eqnarray}
\label{eq.16}
Z=\sqrt{2[(S+B)\text{ln}(1+S/B)-S]}
\end{eqnarray}

In particle physics, the “$5\sigma$” standard is used to define discovery sensitivity, that is, the statistical significance must be $Z>5$; in the case of the “$3\sigma$” standard ($Z>3$), it is claimed that the observed signal has strong evidence. In Table~\ref{tab3}, the 95\% C.L., $3\sigma$ and $5\sigma$ limits of the coefficients $\overline{c}_\gamma$, $\overline{c}_{HB}$, $\overline{c}_{HW}$, $\widetilde{c}_\gamma$, $\widetilde{c}_{HB}$ and $\widetilde{c}_{HW}$ for the center-of-mass energy of 10 TeV with integrated luminosity of 10 ab$^{-1}$ at muon collider are given without and with systematic uncertainty $\delta= 15\%$. The most sensitive limits between all coefficients belong to the coefficients $\overline{c}_{HB}$ and $\overline{c}_{HW}$. In Section~\ref{Sec2}, we mentioned that the coefficients $\overline{c}_{HB}$ and $\overline{c}_{HW}$ make the largest contribution to the total cross-sections as a function of the Wilson coefficients and that these two coefficients are the focus of this study. Table~\ref{tab3} shows that the coefficients $\overline{c}_{HB}$ and $\overline{c}_{HW}$ are the most sensitive limits at the muon collider, and however, the 95\% C.L. limits of these two coefficients are even more sensitive than the experimental studies with the best limits. The limits of the coefficients $\overline{c}_{HB}$ and $\overline{c}_{HW}$ are discussed in detail in Section~\ref{Sec5}, comparing them with the limits of experimental and phenomenological studies.

\begin{table}[H]
\centering
\caption{The limits at 95\% C.L., $3\sigma$ and $5\sigma$ for the coefficients $\overline{c}_\gamma$, $\overline{c}_{HB}$, $\overline{c}_{HW}$, $\widetilde{c}_\gamma$, $\widetilde{c}_{HB}$ and $\widetilde{c}_{HW}$ without and with systematic uncertainty $\delta=15\%$ at muon collider.}
\label{tab3}
\centering
\begin{tabular}{p{2.6cm}p{1.8cm}p{3.8cm}p{3.8cm}p{2.8cm}}
\hline \hline
Coefficients & $\delta_{sys}$ & $\chi^2$ at 95\% C.L. & $Z$ at $3\sigma$ & $Z$ at $5\sigma$     \\ 
\hline
\multirow{2}{*}{$\overline{c}_{\gamma}$} & 0 & [-0.1103; 0.1392] & [-0.2243; 0.2236] & [-0.3052; 0.3027] \\
& 15\% & [-0.2259; 0.2151] & [-0.4465; 0.4478] & [-0.8791; 0.8801] \\
\hline
\multirow{2}{*}{$\widetilde{c}_{\gamma}$} & 0 & [-0.1021; 0.1080] & [-0.2029; 0.2052] & [-0.2773; 0.2724] \\
& 15\% & [-0.2016; 0.2026] & [-0.3742; 0.3819] & [-0.6245; 0.7989] \\
\hline
\multirow{2}{*}{$\overline{c}_{HB}$}
& 0 & [-0.0058; 0.0057] & [-0.0116; 0.0115] & [-0.0162; 0.0161] \\
& 15\% & [-0.0111; 0.0118] & [-0.0255; 0.0254] & [-0.0365; 0.0366] \\ \hline
\multirow{2}{*}{$\widetilde{c}_{HB}$} 
& 0 & [-0.2588; 0.3019] & [-0.5125; 0.4757] & [-0.6661; 0.5970] \\
& 15\% & [-0.5505; 0.4839] & [-1.0174; 0.9116] & [-1.8689; 1.8369] \\ \hline
\multirow{2}{*}{$\overline{c}_{HW}$} 
& 0 & [-0.0047; 0.0035] & [-0.0075; 0.0074] & [-0.0114; 0.0113] \\
& 15\% & [-0.0071; 0.0075] & [-0.0178; 0.0179] & [-0.0251; 0.0252] \\ \hline
\multirow{2}{*}{$\widetilde{c}_{HW}$}
& 0 & [-0.1342; 0.1212] & [-0.2574; 0.2658] & [-0.3420; 0.3741] \\
& 15\% & [-0.2563; 0.2615] & [-0.5186; 0.5250] & [-0.9334; 0.9331] \\ \hline \hline
\end{tabular}
\end{table}

\begin{figure}[H]
\centering
\begin{subfigure}{0.5\linewidth}
\includegraphics[width=\linewidth]{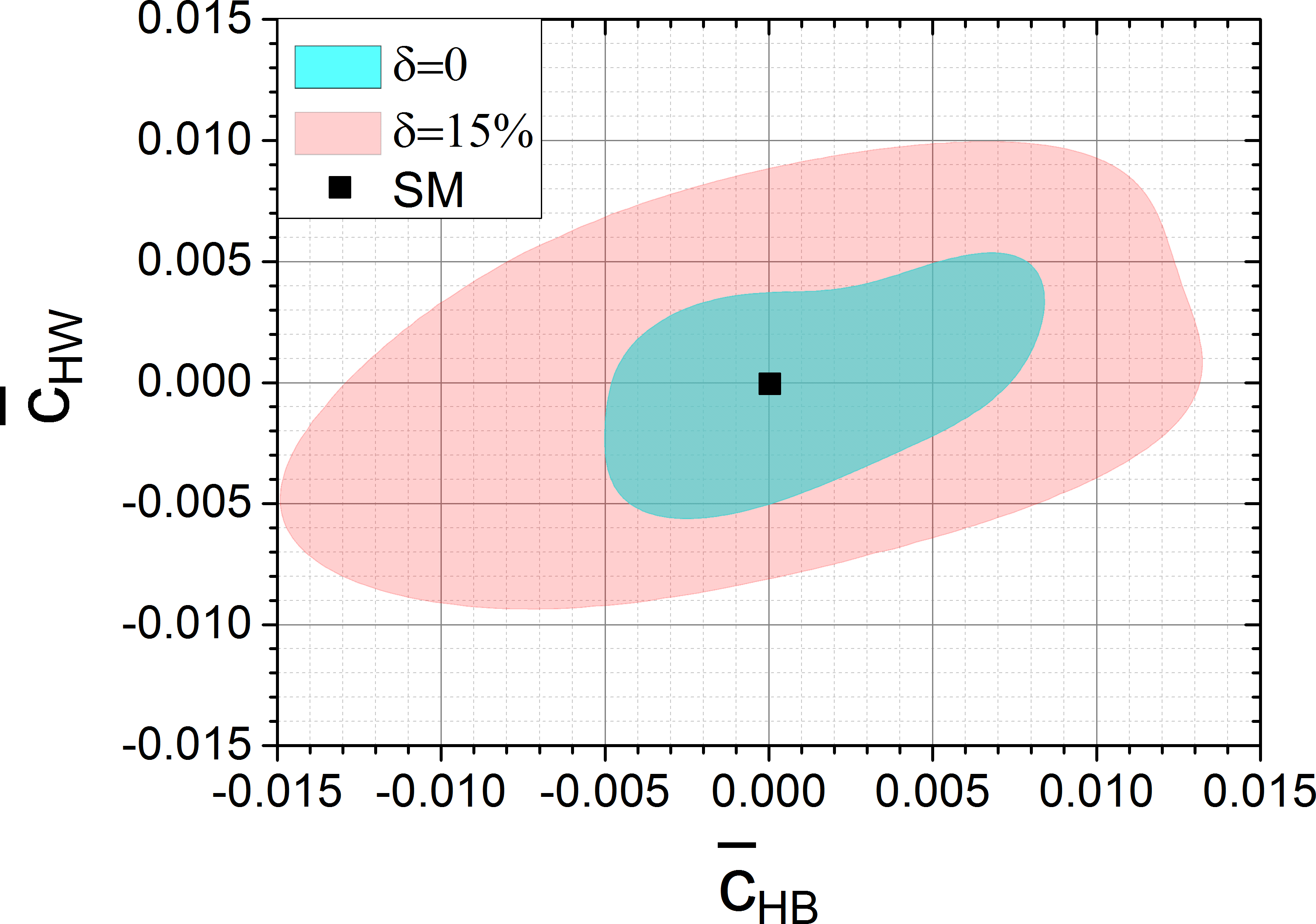}
\caption{}
\label{fig8:a}
\end{subfigure}\hfill
\begin{subfigure}{0.45\linewidth}
\includegraphics[width=\linewidth]{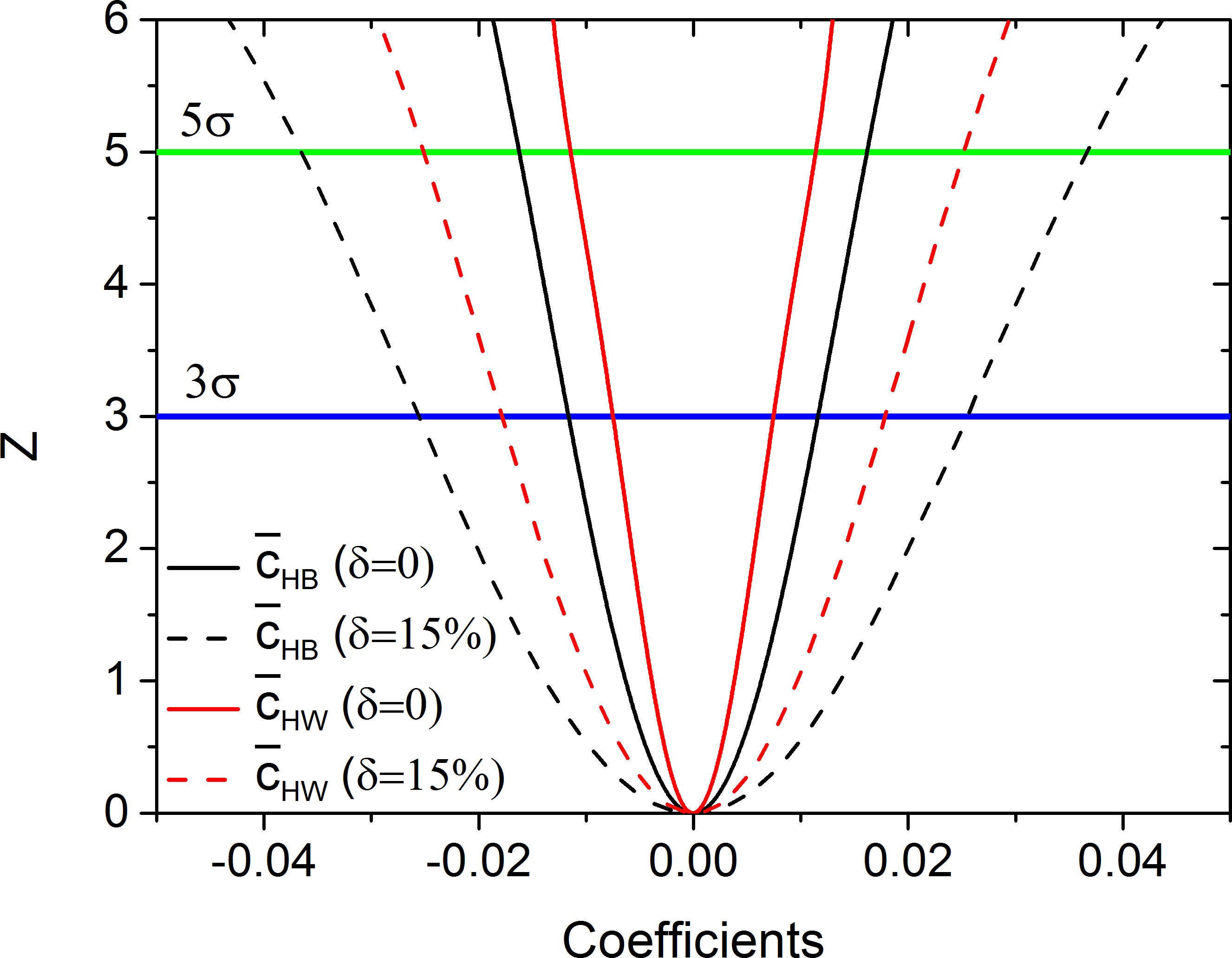}
\caption{}
\label{fig8:b}
\end{subfigure}\hfill
\caption{Two-dimensional 95\% C.L. intervals in plane for $\overline{c}_{HB}-\overline{c}_{HW}$ (left) and the statistical significance ($Z$) as a function of the coefficients $\overline{c}_{HB}$ and $\overline{c}_{HW}$ (right).}
\label{fig8}
\end{figure}

As discussed above, when five of the six Wilson coefficients can be overly constrained, the one-parameter analysis can be beneficial in sensitivity analysis scenarios. To examine the change in the limits of the coefficients $\overline{c}_{HB}$ and $\overline{c}_{HW}$, which is the focus of this study, we consider the limits where all coefficients except the coefficients $\overline{c}_{HB}$ and $\overline{c}_{HW}$ are set to zero. The 95\% C.L. contour is obtained while $\chi^2$, corresponding to two degrees of freedom, is equal to 5.99 in this analysis. The 95\% C.L. contour for the $\overline{c}_{HB}-\overline{c}_{HW}$ plane at the muon collider using the two-parameter analysis is given without and with systematic uncertainty $\delta=15\%$ in Fig.~\ref{fig8:a}.

The statistical significance ($Z$) as a function of the coefficients $\overline{c}_{HB}$ and $\overline{c}_{HW}$ of the Higgs-gauge boson couplings without and with systematic uncertainty $\delta=15\%$ is shown in Fig.~\ref{fig8:b}. From this figure, the bounds on the coefficients $\overline{c}_{HB}$ and $\overline{c}_{HW}$ correspond to the intersection of the curves with the horizontal blue and green lines representing the $3\sigma$ and $5\sigma$ levels, respectively.

The validity of the EFT can be tested by examining the relationship between the new physics scale and the Wilson coefficients of dimension-six operators as $\bar{c} \sim g^2_{*} \nu^2 / \Lambda_s^2$ \cite{Ellis:2015uja,Contino:2013ert,Karadeniz:2020wlo}. Here, $\nu$ is the vacuum expectation value of the Higgs field and $g_*$ is the coupling constant of the heavy degrees of freedom with the SM particles. The upper bound on the new physics scale with $\nu =246$ GeV and $g_{*} = 4\pi$ for strongly-coupled new physics are 40.9 TeV for $\overline{c}_{HB}$:[-0.0058; 0.0057] and 45.1 TeV for $\overline{c}_{HW}$:[-0.0047; 0.0035]. In this study, since the energy scale depends on the specific process and the observable, $E\sim m_{\ell\ell\gamma\gamma}$, these upper bounds are within the validity range of the EFT. An uncertainty arises from the truncation of the EFT expansion because it neglects the contribution from dimension-eight, which may be comparable to or larger than the contribution from dimension-six. However, including the full set of dimension-eight operators is often quite complex; therefore, the SMEFT with only dimension-six effective operators is typically used, but attention should be paid to the potential failure of this truncation and the possible relative contribution of some dimension-eight terms. In this study, the potential truncation error is considered within the systematic uncertainty.

\section{Conclusions} \label{Sec5}

The future muon collider has recently attracted considerable attention due to its multi-TeV energy and clean environment for exploring the Higgs interactions in new physics beyond the SM. At the muon collider with a center-of-mass energy of 10 TeV and an integrated luminosity of 10 ab$^{-1}$, we examine the dimension-six operators of the Higgs-gauge boson couplings $H\gamma\gamma$, $H\gamma Z$ and $HZZ$ through the process $\mu^- \mu^+ \rightarrow Z\gamma\gamma$ using the effective Lagrangian approximation in the SILH basis. We analyze the $\ell \ell \gamma \gamma$ production along with the decay of the $Z$-boson by taking into account the effects of a realistic detector with a tuned muon collider detector. In order to separate the signal events from the backgrounds, we provide the kinematic distributions of the signal and relevant background processes in a cut-based analysis. The 95\% C.L., $3\sigma$ and $5\sigma$ bounds on the Wilson coefficients $\overline{c}_\gamma$, $\overline{c}_{HB}$, $\overline{c}_{HW}$, $\widetilde{c}_\gamma$, $\widetilde{c}_{HB}$ and $\widetilde{c}_{HW}$ in the SILH basis are obtained without and with systematic uncertainty of 15\% using the $\chi^2$ test and statistical significance test from the invariant mass distribution of the $\ell^+ \ell^- \gamma \gamma$ final state.

In this study, we focus on the sensitivity of the coefficients $\overline{c}_{HB}$ and $\overline{c}_{HW}$ at the muon collider. In the experimental study at $\sqrt{s}=13$ TeV with ${\cal L}_{\text{int}}=79.8$ fb$^{-1}$, the ATLAS collaboration determined the 95\% C.L. bound of the coefficient $\overline{c}_{HB}$ and $\overline{c}_{HW}$ in the $H\rightarrow b \overline{b}$ decay channel as [-0.022; 0.049] and [-0.003; 0.008], respectively \cite{Aaboud:2019ygw}. The 95\% C.L. bounds on the Wilson coefficients in the $H\rightarrow \gamma\gamma$ decay channel were investigated by the ATLAS collaboration at $\sqrt{s}=13$ TeV with ${\cal L}_{\text{int}}=139$ fb$^{-1}$, and the experimental limit of the $\overline{c}_{HB}=\overline{c}_{HW}$ coefficients was obtained as [-0.025; 0.022] \cite{ATLAS:2019sdf}. Among the limits of these two experimental studies, the most sensitive limit of the coefficient $\overline{c}_{HB}$ is in Ref.~\cite{ATLAS:2019sdf} and the most sensitive limit of the coefficient $\overline{c}_{HW}$ is in Ref.~\cite{Aaboud:2019ygw}. Comparing the limits of these experimental studies with our limits at 95\% C.L., the limit of the coefficient $\overline{c}_{HB}$ is about 4.1 times more sensitive than in Ref.~\cite{ATLAS:2019sdf}. Our limit at 95\% C.L. of the coefficient $\overline{c}_{HW}$ is about 1.3 times more sensitive than in Ref.~\cite{Aaboud:2019ygw}.

If we consider some phenomenological studies; in Ref.~\cite{Denizli:2018rca}, the 95\% C.L. limits of the $\overline{c}_{HB}$ and $\overline{c}_{HW}$ coefficients at the CLIC with $\sqrt{s}=380$ GeV and ${\cal L}_{\text{int}}=500$ fb$^{-1}$ are [-0.0482; 0.0153] and [-0.00658; 0.00555], respectively, and in Ref.~\cite{Englert:2016tgb}, the 95\% C.L. limits of the $\overline{c}_{HB}$ and $\overline{c}_{HW}$ coefficients at the LHC with $\sqrt{s}=14$ TeV and ${\cal L}_{\text{int}}=3000$ fb$^{-1}$ are [-0.004; 0.004] and [-0.004; 0.004], respectively, in Ref.~\cite{Shi:2019yhw}, the 95\% C.L. limits of $\overline{c}_{HB}$ and $\overline{c}_{HW}$ coefficients at the LHC with $\sqrt{s}=13$ TeV and ${\cal L}_{\text{int}}=36.1$ fb$^{-1}$ are [-0.230; 0.236] and [-0.236; 0.231], respectively, and in Ref.~\cite{Hesari:2018elr}, the 95\% C.L. limits of $\overline{c}_{HW}$ coefficient at the FCC-he with $E_e=60$ GeV and ${\cal L}_{\text{int}}=3$ ab$^{-1}$ are found to be [-0.0032; 0.0035]. According to these phenomenological studies, the limits at 95\% C.L. of the coefficient $\overline{c}_{HB}$ are more sensitive than in Ref.~\cite{Denizli:2018rca} and in Ref.~\cite{Shi:2019yhw} and more stringent than in Ref.~\cite{Englert:2016tgb}. The limits at 95\% C.L. of the coefficient $\overline{c}_{HW}$ are more sensitive than in Ref.~\cite{Denizli:2018rca} and in Ref.~\cite{Shi:2019yhw} and more stringent than in Ref.~\cite{Englert:2016tgb} and Ref.~\cite{Hesari:2018elr}. The comparison of 95\% C.L. limits of the coefficients $\overline{c}_{HB}$ and $\overline{c}_{HW}$ in this study with the limits of present experimental and various phenomenological studies is summarized in Table~\ref{tab4}.

\begin{table}[H]
\centering
\caption{Comparison of the 95\% C.L. limits of $\overline{c}_{HB}$ and $\overline{c}_{HW}$ coefficients in this study with present experimental and phenomenological studies.}
\label{tab4}
\centering
\begin{tabular}{p{4cm}p{3.5cm}p{4cm}p{3.5cm}}
\hline \hline
Coefficients & & $\overline{c}_{HB}$ & $\overline{c}_{HW}$ \\ \hline
This Study & $\chi^{2} \rightarrow 95\%$ C.L. & [-0.0058; 0.0057] & [-0.0047; 0.0035] \\
\hline
Experimental & Ref.~\cite{Aaboud:2019ygw} & [-0.022; 0.049] & [-0.003; 0.008] \\
Limits (ATLAS) & Ref.~\cite{ATLAS:2019sdf} & [-0.025; 0.022] & [-0.025; 0.022] \\
\hline
& Ref.~\cite{Denizli:2018rca} & [-0.0482; 0.0153] & [-0.00658; 0.00555] \\
Phenomenological & Ref.~\cite{Englert:2016tgb} & [-0.004; 0.004] & [-0.004; 0.004] \\
Limits & Ref.~\cite{Shi:2019yhw}  & [-0.230; 0.236] & [-0.236; 0.231] \\
& Ref.~\cite{Hesari:2018elr} & & [-0.0032; 0.0035] \\
   \hline \hline
\end{tabular}
\end{table}

The main motivation for this study is that the clean experimental environment of the future muon collider, combined with its high center-of-mass energy and integrated luminosity, makes it a very promising option for investigating new physics beyond the SM. The more sensitive coefficients $\overline{c}_{HB}$ and $\overline{c}_{HW}$ for the anomalous $H\gamma\gamma$, $HZZ$ and $H\gamma Z$ couplings at the muon collider, compared to the limits of experimental and phenomenological studies, highlight the potential of such colliders. Future muon collider experiments, together with complementary results from hadron colliders, will benefit the investigation of the fundamental nature of Higgs-gauge boson interactions. The results of this study definitely encourage further and more detailed investigations of the multi-TeV muon collider.

\section*{Declaration of competing interest}

The authors declare that they have no known competing financial interests or personal relationships that could have appeared
to influence the work reported in this paper.

\end{document}